\documentclass[twocolumn,prl,twoside,floatfix]{revtex4}

\usepackage{amsmath}
\usepackage{amssymb}
\usepackage{graphicx}
\usepackage{dcolumn}
\usepackage{bm}

\def\d{\mathrm{d}}
\def\epsilon{\varepsilon}
\def\Eq#1{(\ref{#1})}
\def\phi{\varphi}
\def\rho{\varrho}
\def\theta{\vartheta}

\begin{document}


\title{Note on cond-mat/0510119: Jarzynski equation for adiabatically stretched
rotor}

\author{Markus Bier}
\email{bier@fluids.mpi-stuttgart.mpg.de}

\affiliation{
   Max-Planck-Institut f\"ur Metallforschung,  
   Heisenbergstra\ss e 3, 
   70569 Stuttgart, 
   Germany,
}

\affiliation{
   Institut f\"ur Theoretische und Angewandte Physik, 
   Universit\"at Stuttgart, 
   Pfaffenwaldring 57, 
   70569 Stuttgart, 
   Germany
}

\date{October 10, 2005}

\begin{abstract}
In a recent article (cond-mat/0510119) it has been argued that the Jarzynski 
equation is violated for adiabatic stretching processes of a three dimensional
rotor system. Here we want to show that the reasoning is not correct.
Rather, the Jarzynski equation is fulfilled for this adiabatically stretched 
rotor.
\end{abstract}

\maketitle


Consider a particle of mass $\mu$ which can freely move on a sphere around the
origin with temporal varying radius $r(t)$ changing according to some 
prescribed protocol. The Hamiltonian in terms of spherical coordinates 
$\theta,\phi$ and the conjugate momenta $p_\theta,p_\phi$ is
\begin{eqnarray}
   H(\theta,\phi,p_\theta,p_\phi,t) & = & 
   \frac{1}{2\mu r(t)^2}
   \left(p_\theta^2 + \left(\frac{p_\phi}{\sin\theta}\right)^2\right) 
   \nonumber\\ & &
   + U(r(t)),
   \label{eq:hamiltonian}
\end{eqnarray}
where $U$ is an external potential which depends on the sphere radius
$r(t)$ but not on the coordinates $\theta$ or $\phi$. If the system is in 
contact with a heat bath of temperature $\beta^{-1}$ and if the sphere 
radius is constant, $r(t) = R$, a simple calculation yields the canonical
partition function
\begin{equation}
   Z(\beta, R) = \frac{2 \mu R^2}{\beta \hbar^2}\exp(-\beta U(R)).
\end{equation}
Therefore, the free energy difference $\Delta F(\beta,R_1,R_0)$ between the 
equilibrium states for sphere radii $R_1$ and $R_0$ fulfills
\begin{eqnarray}
   & &
   \exp(-\beta\Delta F(\beta,R_1,R_0))
   \nonumber\\
   & = & \frac{Z(\beta,R_1)}{Z(\beta,R_0)}
   \nonumber\\
   & = & 
   \left(\frac{R_1}{R_0}\right)^2\exp(-\beta(U(R_1)-U(R_0))),
   \label{eq:DeltaF}
\end{eqnarray}
which agrees with Ref. \cite{Sung2005}, Eq. (4). 

Now we consider the following stretching process: For $t < 0$, let the system 
be in equilibrium with a heat bath of temperature $\beta^{-1}$ while the 
sphere radius $r(t)$ is held fixed at $R_0$. At $t=0$, the system is decoupled
from the heat bath. In the time interval $t \in [0,t_s]$, the sphere radius
is changed from $r(0) = R_0$ to $r(t_s) = R_1$. At $t=t_s$, the system is again
coupled to the heat bath of temperature $\beta^{-1}$ and the system is 
equilibrated while keeping the sphere radius at $R_1$. After equilibration,
the free energy has changed by $\Delta F(\beta,R_1,R_0)$. On the other 
hand, by changing the sphere radius from $R_0$ to $R_1$, one has to perform 
the work $W$ to the system. In general, $W$ depends on the protocol $r$ and 
the microstate of the system at $t=0$. Remarkably, for a fixed protocol $r$, 
the Jarzynski equation \cite{Jarz1997} asserts 
\begin{equation}
   \langle\exp(-\beta W)\rangle = \exp(-\beta\Delta F(\beta,R_1,R_0)),
   \label{eq:Jarzynski}
\end{equation}
where $\langle\cdot\rangle$ denotes the average over the canonically 
distributed microstates at $t=0$.

It is argued in Ref. \cite{Sung2005} that was ''$W=U(R_1)-U(R_0)$'' which, in
conjunction with Eq. \Eq{eq:DeltaF}, would violate the Jarzynski equation 
\Eq{eq:Jarzynski}. In the following, we want to show that this expression 
for the work $W$ is \emph{not} correct. Rather the Jarzynski equation 
\Eq{eq:Jarzynski} is fulfilled for the system under consideration.   

From Eq. \Eq{eq:hamiltonian} one deduces the Hamiltonian equations
\begin{eqnarray}
   \dot{\theta}   & = & \frac{p_\theta}{\mu r(t)^2} 
   \nonumber\\
   \dot{\phi}     & = & \frac{p_\phi}{\mu r(t)^2 \sin(\theta)^2}
   \nonumber\\
   \dot{p}_\theta & = & \frac{\cos(\theta) p_\phi^2}{\mu r(t)^2 \sin(\theta)^3}
   \nonumber\\
   \dot{p}_\phi   & = & 0.
   \label{eq:hameq}
\end{eqnarray}
The time dependence of the kinetic energy
\begin{equation}
   T =
   \frac{1}{2\mu r(t)^2}
   \left(p_\theta^2 + \left(\frac{p_\phi}{\sin\theta}\right)^2\right) 
\end{equation}
under the Hamiltonian dynamics Eq. \Eq{eq:hameq} is given by
\begin{eqnarray}
   \frac{\d T}{\d t} & = &
   -\frac{\dot{r}(t)}{\mu r(t)^3}
   \left(p_\theta^2 + \left(\frac{p_\phi}{\sin\theta}\right)^2\right) 
   \nonumber\\
   &  &
   + \frac{1}{\mu r(t)^2}
   \left(p_\theta\dot{p_\theta} + \frac{p_\phi\dot{p}_\phi}{\sin(\theta)^2}
   - \frac{\cos(\theta)\dot{\theta}p_\phi^2}{\sin(\theta)^3}
   \right) 
   \nonumber\\
   & = &
   -\frac{\dot{r}(t)}{\mu r(t)^3}
   \left(p_\theta^2 + \left(\frac{p_\phi}{\sin\theta}\right)^2\right) 
   \nonumber\\
   & = &
   -2\frac{\dot{r}(t)}{r(t)}T,
\end{eqnarray}
which yields 
\begin{equation}
   T(r(t)) = \frac{\mathrm{const}}{r(t)^2}.
\end{equation}
The work $W$ performed on the system by stretching the sphere radius of the
rotor from $R_0$ to $R_1$ is given by
\begin{eqnarray}
   W & = & (T(R_1) + U(R_1)) - (T(R_0) + U(R_0)) 
   \nonumber\\
   & = & T(R_0)\left(\left(\frac{R_0}{R_1}\right)^2 - 1\right) + U(R_1)-U(R_0).
\end{eqnarray}
This expression for the work $W$ disagrees with that given in Ref. \cite{Sung2005}. 

For the rotor initially in contact with a heat bath of temperature 
$\beta^{-1}$, $T(R_0)$ is a Boltzmann distributed stochastic variable with 
probability density $\beta\exp(-\beta T(R_0))$. Therefore, one finds
\begin{eqnarray}
   & &
   \langle\exp(-\beta W)\rangle 
   \nonumber\\
   & = &
   \int_0^\infty\d T(R_0)\;\beta\exp(-\beta T(R_0))
   \nonumber\\
   & &
   \exp\left(-\beta T(R_0)\left(\left(\frac{R_0}{R_1}\right)^2 - 1\right) -
   \beta(U(R_1)-U(R_0))\right)
   \nonumber\\
   & = &
   \exp(-\beta(U(R_1)-U(R_0)))
   \nonumber\\
   & &
   \beta\int_0^\infty\d T(R_0)\;
   \exp\left(-\beta\left(\frac{R_0}{R_1}\right)^2T(R_0)\right)
   \nonumber\\
   & = &
   \exp(-\beta(U(R_1)-U(R_0)))\left(\frac{R_1}{R_0}\right)^2.
   \label{eq:W}
\end{eqnarray}

As the rightmost expressions in Eqs. \Eq{eq:DeltaF} and \Eq{eq:W} are 
identical, the validity of the Jarzynski equation \Eq{eq:Jarzynski} for 
the adiabatically stretched rotor is proved.



\end{document}